\documentclass[12pt]{iopart}
\usepackage{graphicx}

\begin{document}

\newcommand{\PrOsSb}{PrOs$_4$Sb$_{12}$}
\newcommand{\LaOsSb}{LaOs$_4$Sb$_{12}$}

\title[Superconductivity and the high field ordered phase in \PrOsSb]
{Superconductivity and the high field ordered phase in the heavy
fermion compound \PrOsSb}

\author{M. B. Maple\dag, P.-C. Ho\dag, N. A. Frederick\dag, V. S. Zapf\dag,
W. M. Yuhasz\dag, E. D. Bauer\dag, A. D. Christianson\ddag, and A. H. Lacerda\ddag}

\address{\dag\ Department of Physics and Institute for Pure and
Applied Physical Sciences, University of California, San Diego, La
Jolla, CA 92093 USA}
\address{\ddag\ National High Magnetic Field Laboratory/LANL, Los Alamos, NM 87545}

\date{\today}

\begin{abstract}

Superconductivity is observed in the filled skutterudite compound
\PrOsSb{} below a critical temperature temperature $T_\mathrm{c} =
1.85$ K and appears to develop out of a nonmagnetic heavy Fermi
liquid with an effective mass $m^{*} \approx 50~m_\mathrm{e}$,
where $m_\mathrm{e}$ is the free electron mass. Features
associated with a cubic crystalline electric field are present in
magnetic susceptibility, specific heat, electrical resistivity,
and inelastic neutron scattering measurements, yielding a
Pr$^{3+}$ energy level scheme consisting of a $\Gamma_{3}$
nonmagnetic doublet ground state, a low lying $\Gamma_{5}$ triplet
excitied state at $\sim 10$ K, and much higher temperature
$\Gamma_{4}$ triplet and $\Gamma_{1}$ singlet excited states.
Measurements also indicate that the superconducting state is
unconventional and consists of two distinct superconducting
phases.  At high fields and low temperatures, an ordered phase of
magnetic or quadrupolar origin is observed, suggesting that the
superconductivity may occur in the vicinity of a magnetic or
quadrupolar quantum critical point.

\end{abstract}


% \maketitle

\section{Introduction}

Since the mid 1990's, several compounds of Pr have been found to
exhibit heavy fermion behavior.  The crystalline electric field
(CEF) ground state of the Pr$^{3+}$ ions in these compounds
appears to be a $\Gamma_{3}$ nonmagnetic doublet that carries an
electric quadrupole moment.  It is conceivable that the heavy
fermion state in these Pr compounds could originate from the
interaction between the Pr$^{3+}$ electric quadrupole moments and
the charges of the conduction electrons.  This would be the
electric analogue of the exchange interaction between the magnetic
dipole moments of Ce or U ions and the conduction electron spins
that is widely believed to be responsible for the heavy fermion
state in most Ce and U heavy fermion compounds.  In fact, such a
mechanism was proposed by Cox in 1987 \cite{Cox87} to account for
the non-Fermi liquid temperature dependences of certain normal
state physical properties of the heavy electron superconductor
UBe$_{13}$.  The Pr compounds that display heavy fermion behavior
include PrInAg$_{2}$ \cite{Yatskar96}, PrFe$_{4}$P$_{12}$
\cite{Sato00}, and, possibly, PrFe$_{4}$Sb$_{12}$ \cite{EBauer02}.

About a year ago, we reported that the compound \PrOsSb{} exhibits
superconductivity  with a superconducting critical temperature
$T_\mathrm{c} = 1.85$ K that apparently develops out of a heavy
Fermi liquid with a quasiparticle effective mass $m^{*} \approx
50~m_\mathrm{e}$, where $m_{e}$ is the mass of the free electron
\cite{Maple01,Bauer02a}. As far as we know,  \PrOsSb{} is the
first example of a heavy fermion  superconductor based on Pr; all
of the other known heavy fermion superconductors are compounds of
Ce or U.  The superconducting state appears to be unconventional
in nature and may consist of two distinct superconducting phases
\cite{Maple02,Izawa02}.   An ordered phase, presumably of magnetic
or quadrupolar origin, occurs at high fields $> 4.5$ tesla and low
temperatures $< 1.5$ K
\cite{Maple02,Ho01,Vollmer02,Oeschler02,Tenya02,Aoki02},
suggesting that the superconductivity may occur in the vicinity of
a magnetic or quadrupolar quantum critical point (QCP). In an
effort to obtain information about the interactions that are
responsible for the heavy fermion state and superconductivity in
\PrOsSb, we have performed measurements of various normal and
superconducting state properties of this compound as a function of
temperature, pressure, and magnetic field
\cite{Maple01,Bauer02a,Maple02,Ho01}. Analysis of magnetic
susceptibility $\chi(T)$, specific heat $C(T)$, electrical
resistivity $\rho(T)$, and inelastic neutron scattering
measurements within the context of a cubic crystalline electric
field (CEF) yields a Pr$^{3+}$ energy level scheme that consists
of a $\Gamma_{3}$ nonmagnetic doublet ground state that carries an
electric quadrupole moment, a low lying $\Gamma_{5}$ triplet
excited state at $\sim 10$ K, and $\Gamma_{4}$ triplet and
$\Gamma_{1}$ singlet excited states at much higher temperatures
($\sim 130$ K and $\sim 313$ K, respectively)
\cite{Maple01,Bauer02a,Maple02,Ho01}.  This scenario suggests that
the underlying mechanism for the heavy fermion behavior in
\PrOsSb{} may involve the interaction of Pr$^{3+}$ electric
quadrupole moments with the charges of the conduction electrons,
rather than Pr$^{3+}$ magnetic dipole moments with the spins of
the conduction electrons. It also raises the possibility that
electric quadrupole fluctuations play a role in the
superconductivity of \PrOsSb. In this paper, we briefly review the
current experimental situation regarding the heavy fermion state,
the superconducting state, and a high field, low temperature phase
that is apparently associated with magnetic or quadrupolar order
in \PrOsSb.

\section{Evidence for a heavy fermion state in \PrOsSb}

\begin{figure}[t]
\begin{center}
\includegraphics[width=8.0cm]{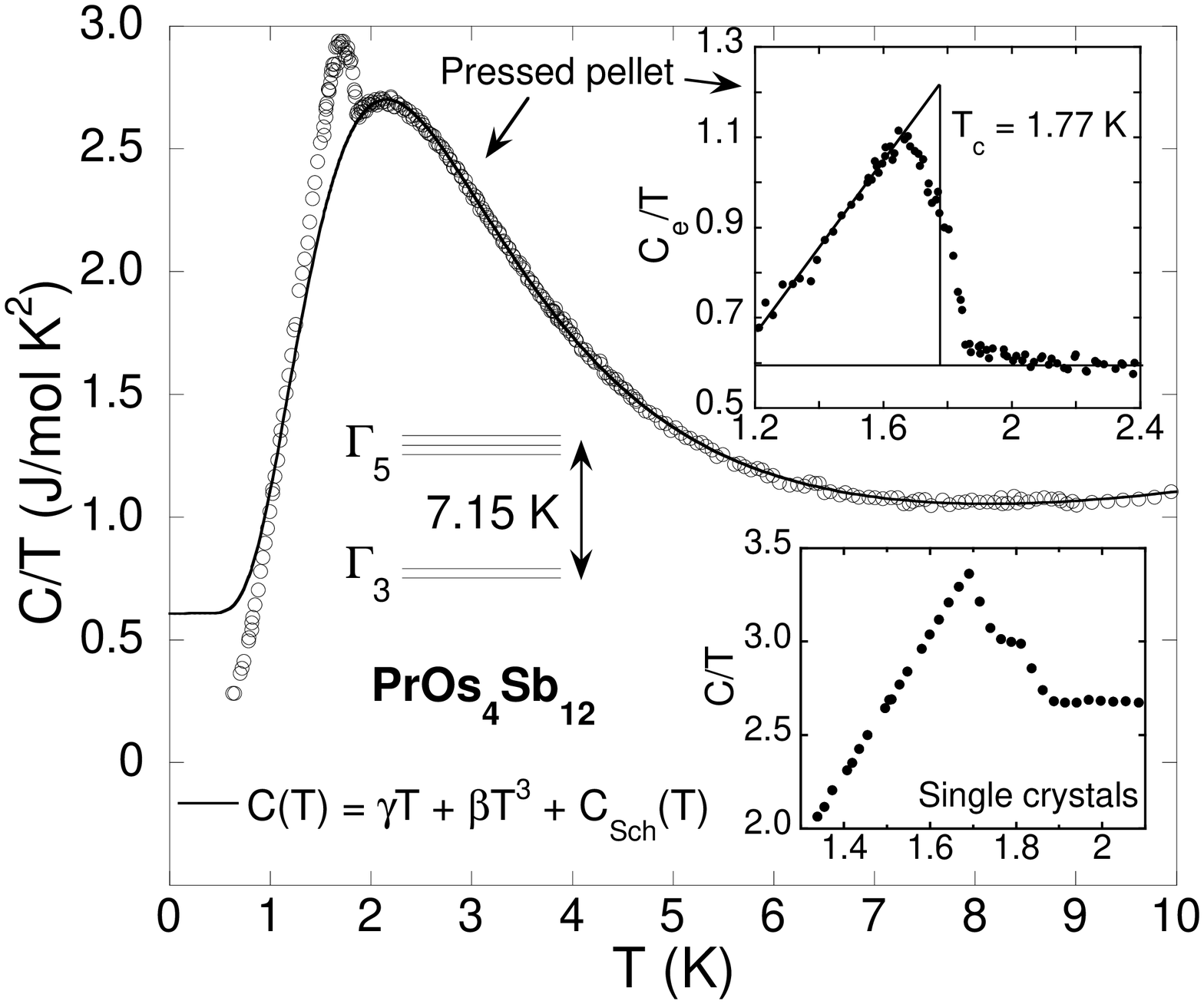}
\end{center}
\caption{Specific heat $C$ divided by temperature $T$, $C/T$, vs
$T$ for a \PrOsSb{} pressed pellet.  The line represents a fit of
the sum of electronic, lattice, and Schottky contributions to the
data.  Upper inset: $C_\mathrm{e}/T$ vs $T$ near $T_\mathrm{c}$
for a \PrOsSb{} pressed pellet ($C_\mathrm{e}$ is the electronic
contribution to $C$).  Lower inset: $C/T$ vs $T$ near
$T_\mathrm{c}$ for \PrOsSb{} single crystals, showing the
structure in $\Delta C$ near $T_\mathrm{c}$. Data from
Refs.~\cite{Maple01,Bauer02a}.} \label{heat}
\end{figure}

The first evidence for a heavy fermion state in the filled
skutterudite compound \PrOsSb{} emerged from specific heat $C(T)$
measurements on a \PrOsSb{} pressed pellet (formed by pressing a
collection of small single crystals in a cylindrical die) at low
temperatures.  Specific heat data  in the form of a plot of $C/T$
vs $T$ between $0.5$ K and $10$ K  for the \PrOsSb{} pressed
pellet from Refs. \cite{Maple01} and \cite{Bauer02a} are shown in
Fig.~\ref{heat}.  The $C(T)$ data have been corrected for excess
Sb derived from the molten Sb flux in which the crystals were
grown.  The line in the figure represents the expression $C(T) =
\gamma T + \beta T^{3} + C_\mathrm{Sch}(T)$, where $\gamma T$ and
$\beta T^{3}$ are electronic and phonon contributions,
respectively, and $C_\mathrm{Sch}(T)$ is a Schottky anomaly for a
two level system consisting of a doublet ground state and a
triplet excited state at an energy $\Delta$ above the ground
state.  The best fit of this expression to the data yields the
values $\gamma = 607$ mJ/mol K$^{2}$, $\beta = 3.95$ mJ/mol
K$^{4}$ (corresponding to a Debye temperature $\theta_\mathrm{D} =
203$ K), and $\Delta = 7.15$ K. Superimposed on the Schottky
anomaly is a feature in the specific heat due to the onset of
superconductivity at $T_\mathrm{c} = 1.85$ K which is also
observed as an abrupt drop in $\rho(T)$ to zero and as a sharp
onset of diamagnetism in $\chi(T)$. The feature in $C(T)/T$ due to
the superconductivity is also shown in the top inset of
Fig.~\ref{heat} along with an entropy conserving construction from
which the ratio of the jump in specific heat $\Delta C$ at
$T_\mathrm{c}$, $\Delta C/T_\mathrm{c} = 632$ mJ/mol K$^{2}$, has
been estimated. Using the BCS relation $\Delta
C/\gamma{}T_\mathrm{c} = 1.43$, we obtain another estimate for
$\gamma$ of $440$ mJ/mol K$^{2}$.  The value of $\Delta
C/T_\mathrm{c}$ is larger than that reported in
Ref.~\cite{Bauer02a} due to the correction of the $C(T)$ data for
the excess Sb (about $30$ percent of the total mass). This value
is comparable to that inferred from the fit to the $C/T$ vs $T$
data in the normal state above $T_\mathrm{c}$, and indicative of
heavy fermion behavior.  A similar analysis of the $C(T)$ data
taken at the University of Karlsruhe on several single crystals of
\PrOsSb{} prepared in our laboratory yielded $\gamma = 313$ mJ/mol
K$^{2}$, $\theta_\mathrm{D} = 165$ K, $\Delta = 7$ K, and $\Delta
C/\gamma{}T_\mathrm{c} \approx 3$, much higher than the BCS value
of $1.43$ and indicative of strong coupling effects
\cite{Vollmer02}. It is interesting that we also find a large
value of $\Delta C/\gamma{}T_\mathrm{c} \approx 3$ in recent
$C(T)$ measurements on one single crystal of \PrOsSb{} at UCSD.
Although the values of $\gamma$ determined from these experiments
vary somewhat, they are all indicative of a heavy electron ground
state and an effective mass $m^{*} \approx 50~m_\mathrm{e}$.

\begin{figure}[t]
\begin{center}
\includegraphics[width=8.0cm]{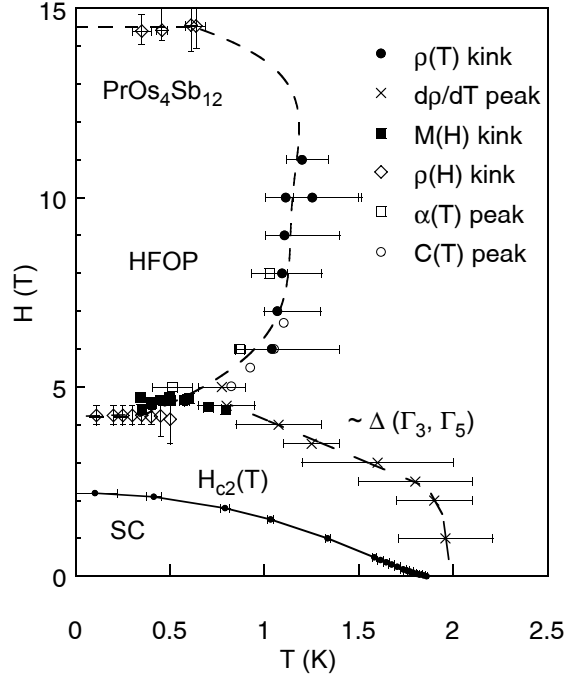}
\end{center}
\caption{Magnetic field - temperature ($H - T$) phase diagram of a
\PrOsSb{} single crystal showing the regions exhibiting
superconductivity (SC) and the high field ordered phase (HFOP).
The boundary delineating the SC phase is based on measurements of
$\rho(H,T)$, while the HFOP phase boundary is based on
$\rho(H,T)$, $C(H,T)$, $M(H,T)$, and $\alpha(H,T)$ measurements
(see text).   The dashed line, derived from a peak in $d\rho/dT$
vs $T$,  is a measure of the splitting between the highest and
lowest Zeeman levels of the Pr$^{3+}~\Gamma_{3}$ ground and
$\Gamma_{5}$ excited states, respectively (see text for further
details). Data from
Refs.~\cite{Bauer02a,Maple02,Ho01,Vollmer02,Oeschler02,Ho03}.}
\label{phase}
\end{figure}

Further evidence of heavy fermion superconductivity is provided by
the upper critical field $H_\mathrm{c2}$ vs $T$ curve shown in
Fig.~\ref{phase} \cite{Bauer02a,Maple02}.  The  orbital critical
field $H^{*}_\mathrm{c2}(0)$ can be derived from the slope  (-19
kOe/K) of the $H_\mathrm{c2}$ curve near $T_\mathrm{c}$ and used
to estimate the superconducting coherence length $\xi_{0} \approx$
\mbox{$116$ \AA{}} via the relation $H^{*}_\mathrm{c2}(0) =
\Phi_{0}/2\pi \xi_{0}^{2}$, where $\Phi_{0}$ is the flux quantum.
The Fermi velocity $v_\mathrm{F}$ can be obtained from the BCS
relation $\xi_{0} = 0.18 \hbar
v_\mathrm{F}/k_\mathrm{B}T_\mathrm{c}$ and used to determine the
effective mass $m^{*}$ by means of the expression $m^{*} = \hbar
k_\mathrm{F}/v_\mathrm{F}$.  Using a simple free electron model to
estimate the Fermi wave vector $k_\mathrm{F}$, an effective mass
$m^{*} \approx 50~m_\mathrm{e}$ is obtained
\cite{Bauer02a,Maple02}. Calculating $\gamma$ from $m^{*}$ yields
$\gamma \sim 350$ mJ/mol K$^{2}$, providing further evidence for a
heavy fermion state in \PrOsSb.

Recently, Sugawara et al.~\cite{Sugawara02} performed de Haas-van
Alphen effect measurements on \PrOsSb.  They found that the
topology of the Fermi surface is close to that of the reference
compound \LaOsSb{} and is explained well by band structure
calculations.  In contrast to the similarity in the Fermi surface
topologies of \PrOsSb{} and \LaOsSb, the cyclotron effective
masses $m_\mathrm{c}^{*}$ of \PrOsSb{} are enhanced up to $\sim 6$
times relative to those of \LaOsSb. The Sommerfeld coefficient
$\gamma$ estimated from the Fermi surface volume and the value of
$m_\mathrm{c}^{*}$, assuming a spherical Fermi surface, is $\sim
150$ mJ/mol K$^{2}$, which is two to three times smaller than the
value of $\gamma$ inferred from the normal and superconducting
properties of \PrOsSb.  Our studies of \LaOsSb{} single crystals
reveal superconductivity with a $T_\mathrm{c}$ of $1$ K.

\section{Normal state of \PrOsSb}

The $\chi(T)$ data for \PrOsSb{} exhibit a peak at $\sim 3$ K and
saturate to a value of $\sim 0.11$ cm$^{3}$/mol as $T \rightarrow
0$, indicative of a nonmagnetic ground state. At temperatures
above $\sim 5$ K, $\chi(T)$ is strongly T-dependent, as expected
for well defined Pr$^{3+}$ magnetic moments.  In the analysis of
the $\chi(T)$ data, interactions between Pr$^{3+}$ ions and
hybridization of the Pr $4$f and conduction electron states were
neglected, while the degeneracy of the Hund's rule multiplet of
the  Pr$^{3+}$ ions was assumed to be lifted by a cubic
crystalline electric field (CEF) and to have a nonmagnetic ground
state.  According to Lea, Leask, and Wolf (LLW) \cite{Lea62}, in a
cubic CEF, the Pr$^{3+} J = 4$ Hund's rule multiplet splits into a
$\Gamma_{1}$ singlet, a $\Gamma_{3}$ nonmagnetic doublet that
carries an electric quadrupole moment, and $\Gamma_{4}$ and
$\Gamma_{5}$ triplets.  It was assumed that the nonmagnetic ground
state of the Pr$^{3+}$ ions corresponds to either a $\Gamma_{1}$
singlet or a $\Gamma_{3}$ nonmagnetic doublet \cite{Bauer02a}.
Although reasonable fits to the $\chi{}(T)$ data could be obtained
for both $\Gamma_{1}$ and $\Gamma_{3}$ ground states, the most
satisfactory fit was obtained for a $\Gamma_{3}$ nonmagnetic
doublet ground state with a $\Gamma_{5}$ first excited triplet
state at $11$ K and $\Gamma_{4}$ and $\Gamma_{1}$ excited states
at $130$ K and $313$ K, respectively. Inelastic neutron scattering
measurements on \PrOsSb{} \cite{Maple02} reveal peaks in the INS
spectrum at $0.71$ meV ($8.2$ K) and $11.5$ meV ($133$ K) that
appear to be associated with transitions between the $\Gamma_{3}$
ground state and the $\Gamma_{5}$ first and $\Gamma_{4}$ second
excited states, respectively, that are in good agreement with the
Pr$^{3+}$ CEF energy level scheme determined from the analysis of
the $\chi(T)$ data.  As noted above, the Schottky anomaly in the
$C(T)$ data on \PrOsSb{} taken at UCSD and at the University of
Karlsruhe \cite{Vollmer02} can be described well by a two level
system consisting of a doublet ground state and a low lying
triplet excited state with a splitting of $\sim 7$ K, a value that
is comparable to the values deduced from the $\chi(T)$ and INS
data.

While a magnetic $\Gamma_{5}$ Pr$^{3+}$ ground state ($\Gamma_{4}$
is not a possible ground state for a cubic system in the LLW
formulation) could also produce a nonmagnetic heavy fermion ground
state via an antiferromagnetic exchange interaction (Kondo
effect), the behavior of $\rho(T)$ of \PrOsSb{} in the normal
state does not resemble the behavior of $\rho(T)$ expected for
this scenario. For a typical magnetically-induced heavy fermion
compound, $\rho(T)$ often increases with decreasing temperature
due to Kondo scattering, reaches a maximum, and then decreases
rapidly with decreasing temperature as the highly correlated heavy
fermion state forms below the coherence temperature.  At low
temperatures, $\rho(T)$ typically varies as $AT^{2}$ with a
prefactor $A \approx 10^{-5}$
[$\mu\Omega$~cm~K$^{2}$(mJ/mol)$^{-2}$] $\gamma^{2}$ that is
consistent with the Kadowaki-Woods relation \cite{Kadowaki86}.  In
contrast,  $\rho(T)$ of \PrOsSb{} \cite{Maple01,Bauer02a} exhibits
typical metallic behavior with negative curvature at higher
temperatures and a pronounced `roll off' below $\sim 8$ K before
it vanishes abruptly when the compound becomes superconducting.
The `roll off' in $\rho{}(T)$ is consistent with a decrease in
charge or spin dependent scattering of conduction electrons by the
Pr$^{3+}$ ions due to the decrease in population of the low lying
first excited state ($\Gamma_{5}$) as the temperature is lowered.
The `roll off' below $\sim 8$ K and the negative curvature at
higher temperatures in $\rho(T)$  can be described reasonably well
by calculations based on magnetic and aspherical Coulomb
scattering of conduction electrons by the Pr$^{3+}$ ions with a
low lying $\Gamma_{5}$ excited state separated from the
$\Gamma_{3}$ ground state by $\sim 6$ K and excited $\Gamma_{4}$
triplet and $\Gamma_{1}$ singlet states at much higher energies
comparable to those found from the analysis of the $\chi(T)$ and
INS measurements \cite{Frederick03}.  The $\rho(T)$ data can be
described by a temperature dependence of the form $AT^{2}$ between
$\sim 8$ K and $45$ K, but with a prefactor $A \approx
0.009~\mu\Omega~$cm/K$^{2}$ that is nearly two orders of magnitude
smaller than that expected from the Kadowaki-Woods relation ($A
\approx 1.2~\mu\Omega$~cm/K$^{2}$ for $\gamma \approx 350$ mJ/mol
K$^{2}$) \cite{Kadowaki86}. Interestingly, $\rho(T)$ is consistent
with $T^{2}$ behavior with a value $A \approx
1~\mu\Omega$~cm/K$^{2}$ in fields of $\sim 5$ tesla \cite{Ho01} in
the high field ordered phase discussed in Section $5$. The
zero-field temperature dependence of $\rho(T)$ is similar to that
observed for the compound PrInAg$_{2}$, which also has a low value
of the coefficient $A$, an enormous $\gamma$ of $\sim 6.5$ J/mol
K$^{2}$, and a $\Gamma_{3}$ nonmagnetic doublet ground state
\cite{Yatskar96}. Another possible source of the enhanced
effective mass in \PrOsSb{} may involve excitations from the
ground state to the the low lying first excited state in the
Pr$^{3+}$ CEF energy level scheme \cite{Fulde02}.

Two studies of the nonlinear magnetic susceptibility have been
performed in an attempt to determine the CEF ground state of the
Pr$^{3+}$ ion in \PrOsSb{} \cite{Tenya02,Bauer02b}. The nonlinear
susceptibility $\chi_{3}$ is the coefficient of the $H^{3}$ term
in the expansion of the magnetization $M$ in a series of odd
powers of $H$; i.e., $M \approx \chi_{1}H + (\chi_{3}/6)H^{3}$,
where $\chi_{1}$ is the ordinary linear susceptibility.  In an
ionic situation, $\chi_{3}$ is isotropic and varies as $T^{-3}$
for a magnetic ground state, whereas $\chi_{3}$ is anisotropic and
diverges at low temperatures for $H~||~[100]$ and approaches a
constant for $H~||~[111]$ for a non-Kramers $\Gamma_{3}$ doublet
ground state \cite{Morin81}.  This type of study was previously
employed in an attempt to determine the ground state of U in the
compound UBe$_{13}$ \cite{Ramirez94}. In both studies of
\PrOsSb{}, $\chi_{3}(T)$ was found to behave similarly for
$H~||~[100]$ and $H~||~[111]$, exhibiting a minimum near $4$ K
followed by a maximum near $1$ K and a negative divergence with
decreasing temperature. Calculations based on the quadrupolar
Anderson-Hamiltonian described the $\chi_{3}(T)$ data reasonably
well for $H~||~[100]$, but not very well for $H~||~[111]$. It was
concluded that the data were qualitatively consistent with a
$\Gamma_{3}$ ground state, given the limitations of the experiment
and the complexity of the theory.  The $\chi_{3}(T)$ studies are
difficult to interpret because of the curvature of $M(H)$ and the
complications that arise at lower temperatures $T \leq
T_\mathrm{c}$ and lower fields $H \leq H_\mathrm{c2}$ due to the
superconductivity and at temperatures $T \leq 2$ K and higher
fields $H \geq 4.5$ tesla by the onset of the high field ordered
phase, discussed in Section $5$.

\section{Superconducting state of \PrOsSb}

A number of features in the superconducting properties of
\PrOsSb{} indicate that the superconductivity of this compound is
unconventional in nature.  One of these features is the
`double-step' structure in the jump in $C(T)$ near $T_\mathrm{c}$
in single crystals (lower inset of Fig.~\ref{heat}) that suggests
two distinct superconducting phases with different
$T_\mathrm{c}$'s: $T_\mathrm{c1} \approx 1.85$ K and
$T_\mathrm{c2} \approx 1.70$ K \cite{Maple02,Vollmer02}.  This
structure is not evident in the $C(T)$ data taken on the pressed
pellet of \PrOsSb{} shown in the upper inset of Fig.~\ref{heat},
possibly due to strains in the single crystals out of which the
pressed pellet is comprised that broaden the transitions at
$T_\mathrm{c1}$ and $T_\mathrm{c2}$ so that they overlap and
become indistinguishable. However, at this point, we are unable to
eliminate the possibility  that the two apparent jumps in $C(T)$
are due to sample inhomogeneity.  It is noteworthy that all of the
single crystal specimens prepared in our laboratory and
investigated by our group and our collaborators exhibit this
`double-step' structure. Multiple superconducting transitions,
apparently associated with distinct superconducting phases, have
previously been observed in two other heavy fermion
superconductors, UPt$_{3}$ \cite{Lohneysen94} and
U$_{1-x}$Th$_{x}$Be$_{13}~(0.1 \leq x \leq 0.35)$ \cite{Ott85}.
Measurements of the specific heat in magnetic fields reveal that
the two superconducting features shift downward in temperature at
nearly the same rate with increasing field, consistent with the
smooth temperature dependence of the $H_\mathrm{c2}(T)$ curve
\cite{Vollmer02}. These two transitions have also been observed in
thermal expansion measurements \cite{Oeschler02}, which, from the
Ehrenfest relation, reveal that $T_\mathrm{c1}$ and
$T_\mathrm{c2}$ have different pressure dependences, suggesting
that they are associated with two distinct superconducting phases.
Another feature is the power law T-dependence of
$C_\mathrm{s}(T)$, $C_\mathrm{s}(T) \sim T^{2.5}$, after the
Schottky anomaly and $\beta T^{3}$ lattice contributions have been
subtracted from the $C(T)$ data. (As reported in
Ref.~\cite{Maple02}, $C_\mathrm{s}(T)$ follows a power law with
$C_\mathrm{s}(T) \sim T^{3.9}$ when the Schottky anomaly is not
subtracted.)  However, this dependence can only be established
from $T_\mathrm{c}$ down to $\sim 0.4~T_\mathrm{c}$, since it is
not possible to reliably correct the $C(T)$ data at lower
temperatures for an enormous nuclear Schottky anomaly.  Power law
T-dependences of the superconducting properties are generally
attributed to nodes in the superconducting energy gap at points or
lines on the Fermi surface.  Among three recent experiments on
\PrOsSb, described below, two yield evidence for an isotropic
energy gap, while another provides evidence for two distinct
superconducting phases in the H-T plane with different numbers of
point nodes in the energy gap.

Recent transverse field $\mu$SR \cite{MacLaughlin02} and Sb-NQR
measurements \cite{Kotegawa02} on \PrOsSb{} are consistent with an
isotropic energy gap. Along with the specific heat, these
measurements indicate strong coupling superconductivity. These
findings suggest an s-wave, or, perhaps, a Balian-Werthamer p-wave
order parameter. On the other hand, the superconducting gap
structure of \PrOsSb{} was investigated by means of thermal
conductivity measurements in magnetic fields rotated relative to
the crystallographic axes by Izawa et al. \cite{Izawa02}.  These
measurements reveal two regions in the $H-T$ plane, a low field
region in which $\Delta({\bf k})$ has two point nodes, and a high
field region where $\Delta({\bf k})$ has six point nodes.  The
line lying between the low and high field superconducting phases
may be associated with the transition at $T_\mathrm{c2}$, whereas
the line between the high field superconducting phase and the
normal phase, $H_\mathrm{c2}(T)$, converges with $T_\mathrm{c1}$
as $H \rightarrow 0$.  Clearly, more research will be required to
further elucidate the nature of the superconductivity in \PrOsSb.

\begin{figure}[t]
\begin{center}
\includegraphics[width=8.0cm]{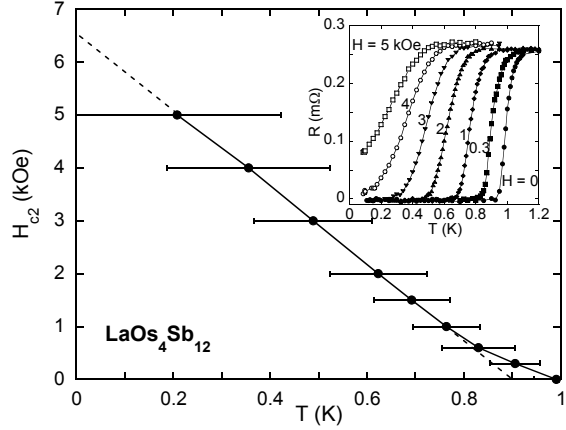}
\end{center}
\caption{Upper critical field $H_\mathrm{c2}$ vs $T$ for \LaOsSb,
based on resistance $R$ vs temperature $T$ data shown in the
inset.} \label{uppercrit}
\end{figure}

It is noteworthy that the reference compound without 4f electrons,
\LaOsSb, is also superconducting, but with $T_\mathrm{c}$
$\approx$ 1 K, considerably smaller than that of \PrOsSb.  This
suggests that the pairing interaction is enhanced by the presence
of the Pr $4$f electrons, possibly through the interaction of the
Pr$^{3+}$ electric quadrupole moments with the conduction
electrons.  In contrast, the values of $T_\mathrm{c}$ of the other
PrT$_{4}$X$_{12}$ filled skutterudites are smaller than their
La-based counterparts. The values of $T_\mathrm{c}$ in K for
MT$_{4}$X$_{12}$ compounds with M = La or Pr are listed in
Table~\ref{Tc}. A plot of the upper critical field $H_\mathrm{c2}$
vs $T$ for \LaOsSb{} and the resistive $R(T)$ transition curves
upon which it is based are shown in Fig.~\ref{uppercrit}.

\begin{table}[t]
\caption{Superconducting critical temperature $T_\mathrm{c}$ of
LnT$_{4}$X$_{12}$ compounds for Ln = La, Pr; T = Fe, Ru, Os; and X
= P, As, Sb. The $T_\mathrm{c}$ values listed are in K and have
been derived from references
\cite{EBauer02,Kotegawa02,Meisner81,Shirotani00,Ravot01,Uchiumi99,
Torikachvili87,Sekine97,Takeda00,Shirotani97}. The symbols \dag{}
and \ddag{} indicate that superconductivity has not been observed
in those compounds above $2$ K and $0.35$ K, respectively.}
\begin{minipage}{\linewidth}\begin{indented}\item{}
\vspace{1cm}
\begin{tabular}{c|c|c|c|c|c|c|c|}
 \multicolumn{1}{c}{} & \multicolumn{3}{c}{LaT$_{4}$X$_{12}$}
 & \multicolumn{1}{c}{} & \multicolumn{3}{c}{PrT$_{4}$X$_{12}$} \\
 \cline{2-4}\cline{6-8}
 Fe & 4.1 & \dag & \dag & & \ddag & ? & \dag \\ \cline{2-4}\cline{6-8}
 Ru & 7.2 & 10.3 & 2.8 & & \ddag & 2.4 & 1.0 \\ \cline{2-4}\cline{6-8}
 Os & 1.8 & 3.2 & 1.0 & & \dag & ? & 1.8 \\ \cline{2-4}\cline{6-8}
 \multicolumn{1}{c}{} & \multicolumn{1}{c}{P} & \multicolumn{1}{c}{As}
 & \multicolumn{1}{c}{Sb} & \multicolumn{1}{c}{} & \multicolumn{1}{c}{P}
 & \multicolumn{1}{c}{As} & \multicolumn{1}{c}{Sb} \\
\end{tabular}
\end{indented}\end{minipage}
\label{Tc}
\end{table}

\section{High field ordered phase in \PrOsSb}

\begin{figure}[t]
\begin{center}
\includegraphics[width=8.0cm]{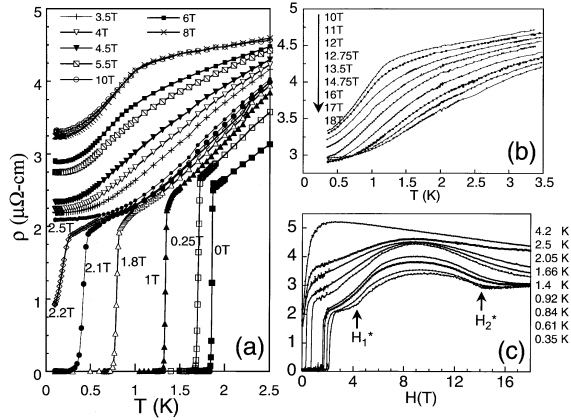}
\end{center}
\caption{(a) and (b): Electrical resistivity $\rho$ vs temperature
$T$ at various magnetic fields $H$ up to 18 tesla (T) for a
\PrOsSb{} single crystal.  (c) $\rho$ vs $H$ at various
temperatures between $0.35$ K and $4.2$ K .  The rapid drop in
$\rho$ to zero for $H < 2.5$ T is due to the superconducting
transition, while the shoulder in $\rho(T)$ at $\sim 1$ K above
$4.5$ T and the kinks in $\rho(H)$ (indicated at $H_{1}^{*}$ and
$H_{2}^{*}$) below $1.7$ K are due to a field induced phase (high
field ordered phase - HFOP). After Refs.~\cite{Ho01,Ho03}.}
 \label{magrho}
\end{figure}

Evidence for a high field ordered phase was first derived from
magnetoresistance measurements in the temperature range $80$ mK
$\leq T \leq 2$ K and magnetic fields up to $10$ tesla
\cite{Maple02,Ho01}. Recently, the magnetoresistance measurements
have been extended up to $18$ tesla for $0.35$ K $\leq T \leq 2$ K
\cite{Ho03}. The $H-T$ phase diagram, depicting the
superconducting region and the high field ordered phase, is shown
in Fig.  2.  The line that intersects the high field ordered phase
represents the inflection point of the `roll-off' in $\rho(T)$ for
$H < 4.5$ tesla at low temperatures and is a measure of the
splitting between the Pr$^{3+}$ ground state and the first excited
state, which decreases with field (see Fig.~\ref{magrho}).  The
high field ordered phase has also been observed by means of large
peaks in the specific heat \cite{Vollmer02,Aoki02} and thermal
expansion \cite{Oeschler02} and kinks in magnetization vs magnetic
field curves \cite{Tenya02,Ho03} in magnetic fields $> 4.5$ tesla
and temperatures $< 1.5$ K.

Shown in Fig.~\ref{magrho} (a) and (b) are $\rho(T)$ data for
various magnetic fields up to $18$ tesla for \PrOsSb, which reveal
drops in $\rho(T)$ due to the superconductivity for $H \leq 2.2$
tesla and features in $\rho(T)$ associated with the onset of the
high field ordered phase for $H \geq 4.5$ tesla.  Isotherms of
electrical resistivity $\rho$ vs $H$ at various temperatures
$0.35$ K $\leq T \leq 4.2$ K and fields $0 \leq H \leq 18$ tesla
are shown in Fig.~\ref{magrho}(c). The fields denoting the
boundaries of the high field ordered phase, $H_{1}^{*}$ and
$H_{2}^{*}$, are indicated in the figure.

\section{Summary}

Experiments on the filled skutterudite compound \PrOsSb{} have
revealed a number of extraordinary phenomena: a heavy fermion
state characterized by an effective mass $m^{*} \approx
50~m_\mathrm{e}$, unconventional superconductivity below
$T_\mathrm{c} = 1.85$ K with two distinct superconducting phases,
and a high field ordered phase, presumably associated with
magnetic or quadrupolar order. Analysis of $\chi(T)$, $C(T)$,
$\rho(T)$, and INS data indicate that Pr$^{3+}$ has a nonmagnetic
$\Gamma_{3}$ doublet ground state that carries an electric
quadrupole moment, a low lying $\Gamma_{5}$ triplet excited state
at $\sim 10$ K, and $\Gamma_{4}$ triplet and $\Gamma_{1}$ singlet
excited states at much higher energies.  This suggests that the
interaction between the quadrupole moments of the Pr$^{3+}$ ions
and the charges of the conduction electrons, as well as the
excitations between the $\Gamma_{3}$ ground state and $\Gamma_{5}$
low lying excited state, may play an important role in generating
the heavy fermion state and superconductivity in this compound.
The heavy fermion state, unconventional superconductivity, and
high field ordered phase observed in \PrOsSb{} and reviewed herein
present a significant challenge for theoretical description
\cite{Miyake02,Maki02,Goryo02,Anders02,Norman94}.

This research was supported by the US Department of Energy (DOE),
Grant No. DE-FG03-86ER-45230; the US National Science Foundation
(NSF), Grant No. DMR-00-72125; and the NEDO International Joint
Research Program at UCSD. Los Alamos National Laboratory is
supported by the NSF, the State of Florida, and the US DOE.

%%%%%%%%%%%%%%%%%%%%%%%%%% Figure %%%%%%%%%%%%%%%%%%%%%%%%%%%%%%%%%%%%%
%goes in main text area at location desired
%\begin{figure}[t]
%\begin{center}
%\includegraphics[width=8.0cm]{sample-fig.eps}
%\end{center}
%\caption{This is a sample of figure.}
%\end{figure}
%%%%%%%%%%%%%%%%%%%%%%%%%%%%%%%%%%%%%%%%%%%%%%%%%%%%%%%%%%%%%%%%%%%%%%%


\section*{References}

\begin{thebibliography}{99}

\bibitem{Cox87} Cox D L 1987 {\it Phys. Rev. Lett.} {\bf 59} 1240.

\bibitem{Yatskar96} Yatskar A, Beyermann W P, Movshovich R, and
Canfield P C 1996 {\it Phys. Rev. Lett.} {\bf 77} 3637.

\bibitem{Sato00} Sato H, Abe Y, Okada H, Matsuda T D,
Abe K, Sugawara K, and Aoki Y 2000 {\it Phys. Rev. B} {\bf 62}
15125.

\bibitem{EBauer02} Bauer E, Berger St, Galatanu A, Paul Ch, Della Mea M,
Michor H, Hilscher G, Grytsiv A, Rogl P, Kaczorowski D, Keller L,
Hermannsd\"{o}rfer T, and Fischer P 2002 {\it Physica B} {\bf
312-313} 840.

\bibitem{Maple01} Maple M B, Bauer E D, Zapf V S, Freeman E J,
Frederick N A, and Dickey R P 2001 {\it Acta Physica Polonica B}
{\bf 32} 3291.

\bibitem{Bauer02a} Bauer E D, Frederick N A, Ho P-C,
Zapf V S, and Maple M B 2002 {\it Phys. Rev. B} {\bf 65} 100506
(R).

\bibitem{Maple02} Maple M B, Ho P-C, Zapf V S,
Frederick N A, Bauer E D, Yuhasz W M, Woodward F M, and Lynn J W
2002 {\it J. Phys. Soc. Jpn.} {\bf 71} Suppl. 23.

\bibitem{Izawa02} Izawa K, Nakajima Y, Goryo J, Matsuda Y,
Osaki S, Sugawara H, Sato H, Thalmeier P, and Maki K 2002
cond-mat/0209553.

\bibitem{Ho01} Ho P-C, Zapf V S, Bauer E D,
Frederick N A, Maple M B, Geister G, Rogl P, Berger St, Paul Ch,
and Bauer E 2001 in {\it Physical Phenomena at High Magnetic
Fields - IV} Boebinger G, Fisk Z, Gor'kov L P, Lacerda A, and
Schrieffer J R, eds. (World Scientific, Singapore) 98-103.

\bibitem{Vollmer02} Vollmer R, Fai$\beta$t A, Pfleiderer C,
von L\"{o}hneysen H, Bauer E D, Ho P-C, and Maple M B 2002
cond-mat/0207225.

\bibitem{Oeschler02} Oeschler N, Gegenwart P, Steglich F,
Frederick N A, Bauer E D, and Maple M B 2002 {\it Acta Physica
Polonica B} (Proc. SCES 2002, Krakow, Poland) to be published.

\bibitem{Tenya02} Tenya K, Oeschler N, Gegenwart P,
Steglich F, Frederick N A, Bauer E D, and Maple M B 2002 in {\it
LT23} to be published.

\bibitem{Aoki02} Aoki Y, Namiki T, Ohsaki S, Saha S R,
Sugawara H, and Sato H 2002 {\it J. Phys. Soc. Jpn.} {\bf 71}
2098.

\bibitem{Sugawara02} Sugawara H, Osaki S, Saha S R,
Aoki Y, Sato H, Inada Y, Shishido H, Settai R, Onuki Y, Harima H,
and Oikawa K 2002 {\it Phys. Rev. B} {\bf 66} 220504(R).

\bibitem{Lea62} Lea K R, Leask M J M, and Wolf W P 1962
{\it J. Phys. Chem. Solids} {\bf 23} 1381.

\bibitem{Kadowaki86} Kadowaki K and Woods S B 1986 {\it Solid State
Commun.} {\bf 58} 507.

\bibitem{Frederick03} Frederick N A and Maple M B 2003 unpublished.

\bibitem{Fulde02} Fulde P 1997 {\it Physica B} {\bf 230-232}
1.

\bibitem{Bauer02b} Bauer E D, Ho P-C, Maple M B,
Schauerte T, Cox D L, and Anders F B 2002 {\it Phys. Rev. B}
submitted.

\bibitem{Morin81} Morin P and Schmitt D 1981 {\it Phys. Rev. B}
{\bf 23} 5936.

\bibitem{Ramirez94} Ramirez A P, Chandra P, Coleman P,
Fisk Z, Smith J L, and Ott H R 1994 {\it Phys. Rev. Lett.} {\bf
73} 3018.

\bibitem{Lohneysen94} von L\"{o}hneysen H 1994 {\it Physica B}
{\bf 197} 551.

\bibitem{Ott85} Ott H R, Rudiger H, Fisk Z, and Smith J L 1985
{\it Phys. Rev. B} {\bf 31} 1651.

\bibitem{MacLaughlin02} MacLaughlin D E, Sonier J E,
Heffner R H, Bernal O O, Young B L, Rose M S, Morris G D, Bauer E
D , Do T D, and Maple M B 2002 {\it Phys. Rev. Lett} {\bf 89}
157001-1.

\bibitem{Kotegawa02} Kotegawa H, Yogi M, Imamura Y,
Kawasaki Y, Zheng G-Q, Kitaoka Y, Ohsaki S, Sugawara H, Aoki Y,
and Sato H 2002 cond-mat/0209106.

\bibitem{Ho03} Ho P-C, Frederick N A, Zapf V S, Bauer E D, Do T D,
Maple M B, Christianson A D, and Lacerda A H, to be published.

\bibitem{Miyake02} Miyake K, Kohno H, and Harima H 2002
in {\it LT23} to be published.

\bibitem{Meisner81} Meisner G P 1981 {\it Physica B} {\bf 108} 763.

\bibitem{Shirotani00} Shirotani I, Ohno K, Sekine C, Yagi T, Kawakami T,
Nakanishi T, Takahashi H, Tang J, Matsushita A, and Matsumoto T
2000 {\it Physica B} {\bf 281 \& 282} 1021.

\bibitem{Ravot01} Ravot D, Lafont U, Chapon L, Tedenac J C, and Mauger A 2001
{\it J. Alloys Compd.} {\bf 323-324} 389.

\bibitem{Uchiumi99} Uchiumi T, Shirotani I, Sekine C, Todo S, Yagi T,
Nakazawa Y, and Kanoda K 1999 {\it J. Phys. Chem. Solids} {\bf 60}
689.

\bibitem{Torikachvili87} Torikachvili M S, Chen J W, Dalichaouch Y,
Guertin R P, McElfresh M W, Rossel C, Maple M B, and Meisner G P
1987 {\it Phys. Rev. B} {\bf 36} 8660.

\bibitem{Sekine97} Sekine C, Uchiumi T, Shirotani I, and Yagi T 1997
{\it Phys. Rev. Lett.} {\bf 79} 3218.

\bibitem{Takeda00} Takeda N and Ishikawa M 2000 {\it J. Phys. Soc. Jpn.}
{\bf 69} 868.

\bibitem{Shirotani97} Shirotani I, Uchiumi T, Ohno K, Sekine C, Nakazawa Y,
Kanoda K, Todo S, and Yagi T 1997 {\it Phys. Rev. B} {\bf 56}
7866.

\bibitem{Maki02} Maki K, Thalmeier P, Yuan Q, Izawa K, and Matsuda Y 2002
cond-mat/0212090.

\bibitem{Goryo02} Goryo J 2002 cond-mat/0212094.

\bibitem{Anders02} Anders F B 2002 {\it Phys. Rev. B} {\bf 66}
020504(R).

\bibitem{Norman94} Norman M R 1994 {\it Phys. Rev. Lett.} {\bf 72} 2077.




\end{thebibliography}
\end{document}